\documentclass[11pt]{article}

\usepackage[T1]{fontenc}
\usepackage{tgpagella}
\usepackage{mathpazo}

\usepackage[paperheight=11in,paperwidth=8.5in,margin=1in]{geometry}

\usepackage{microtype}

\usepackage{amsmath}
\usepackage{graphicx}
\usepackage{booktabs}
\usepackage{array}
\usepackage{longtable}

\usepackage[affil-it]{authblk}
\usepackage{hyperref}

\usepackage[table]{xcolor}
\usepackage{booktabs}
\usepackage{pgfplots}
\pgfplotsset{compat=1.18}

\usepackage{xcolor}
\definecolor{darkblue}{rgb}{0, 0, 0.5}
\usepackage{hyperref}
\hypersetup{colorlinks=true, citecolor=darkblue, linkcolor=darkblue, urlcolor=darkblue}
\usepackage{pgfplots}
\pgfplotsset{compat=1.18}

\usepackage{listings}
\usepackage{xcolor}

\lstdefinestyle{promptbox}{
    backgroundcolor=\color{gray!5},
    basicstyle=\ttfamily\footnotesize,
    breaklines=true,
    captionpos=b,
    lineskip=-0.5pt,
    commentstyle=\color{gray},
    frame=single,
    framerule=0.5pt,
    rulecolor=\color{gray!30},
    keepspaces=true,
    showspaces=false,
    showstringspaces=false,
    xleftmargin=10pt,
    xrightmargin=10pt,
    framesep=10pt,
}

\renewenvironment{abstract}{\vskip.075in\centerline{\large\bf Abstract}\vspace{0.5ex}\begin{quote}}{\par\end{quote}\vskip 1ex}

\title{Technical Report: Exploring the Emerging Threats of the Agent Skill Ecosystem}

\date{February 5, 2026}

\author[1]{Luca Beurer-Kellner\thanks{Alphabetical author ordering. Corresponding author: \href{mailto:luca.beurer-kellner@snyk.io}{luca.beurer-kellner@snyk.io}}}
\author[1]{Aleksei Kudrinskii}
\author[1]{Marco Milanta}
\author[1]{Kristian Bonde Nielsen}
\author[1]{Hemang Sarkar}
\author[1]{Liran Tal}

\affil[1]{Snyk}

\usepackage[table]{xcolor}

\definecolor{riskCritical}{HTML}{CC0000} 
\definecolor{riskHigh}{HTML}{FF8C00}     
\definecolor{riskMedium}{HTML}{EAB308}   

\newcommand{\badge}[2]{%
  \begingroup
  \sffamily\scriptsize\bfseries
  \setlength{\fboxsep}{2pt}%
  \colorbox{#1}{\color{white}#2}%
  \endgroup
}

\begin{document}

\maketitle

\begin{abstract}
We analyzed 3,984 AI agent skills from major marketplaces and found 76 confirmed malicious payloads, including credential theft, backdoor installation, and data exfiltration. 13.4\% of all skills contain at least one critical-level security issue and at least 8 manually confirmed malicious skills remain publicly available on clawhub.ai as of the date of publication. This report documents our methodology, presents a threat taxonomy based on real-world samples, and details the attack patterns we observed. As skill marketplaces grow rapidly and AI agents gain access to sensitive credentials and systems, automated security analysis is no longer optional.
\end{abstract}

\section{Introduction}

The AI agent ecosystem is experiencing unprecedented growth. As of early 2026, major skill marketplaces like \href{https://clawhub.ai/skills}{ClawHub} and \href{http://skills.sh}{skills.sh} host thousands of ``skills'', packaged instructions and code that extend AI agents' capabilities\footnote{Agent Skill Specification, \url{https://agentskills.io}}. These skills enable agents to perform new tasks ranging from library-specific code generation and data analysis to trading automation and system administration.

\begin{figure}[h]
\centering
\includegraphics[width=1.0\linewidth]{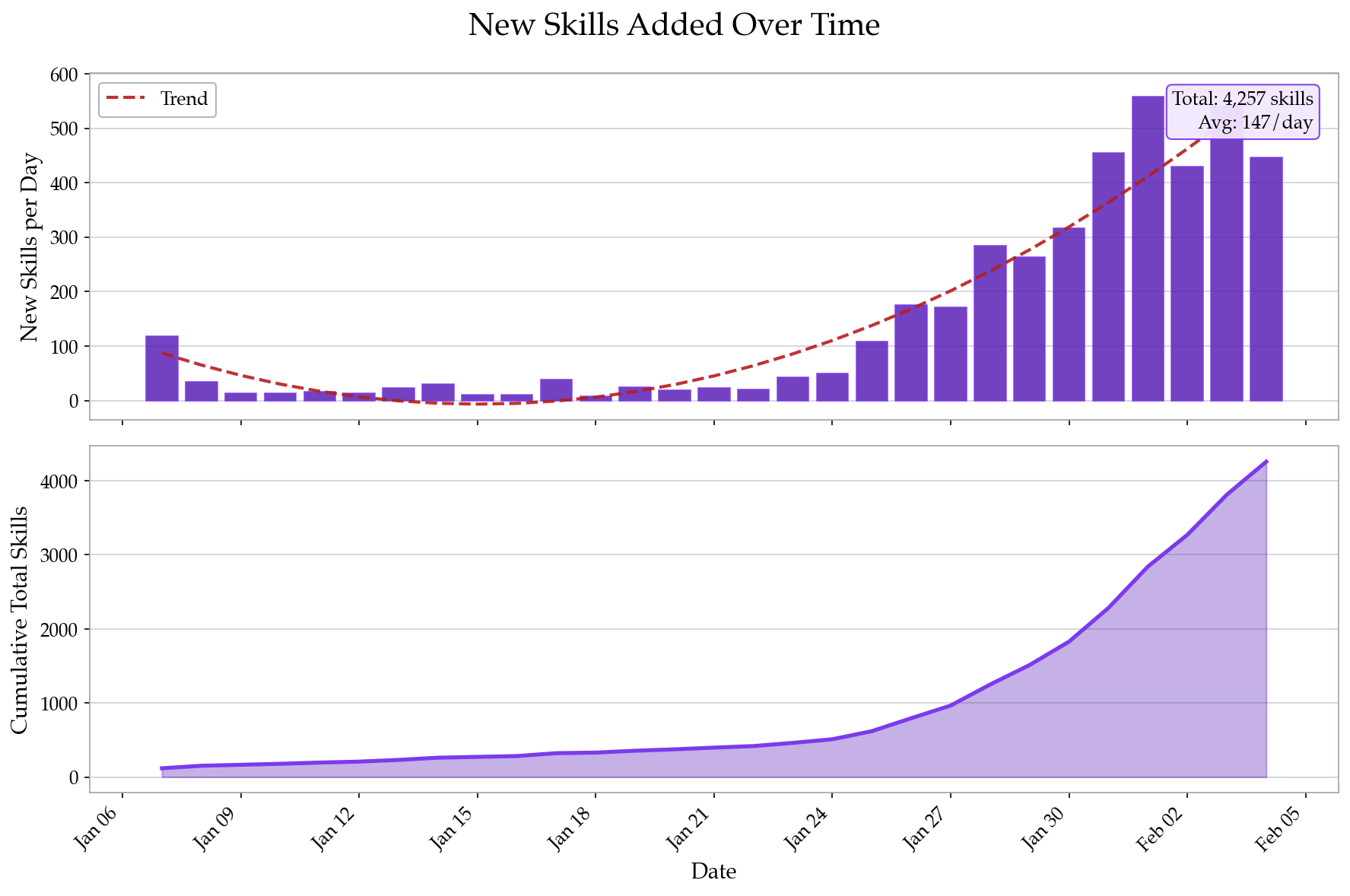}
\caption{Number of agent skills published every day throughout 2026.}
\end{figure}

\textbf{Malicious Skills Emerge} Researchers have previously demonstrated the theoretical possibility of skill-based attacks \cite{schmotz2025agent}. However, in February 2026, OpenSourceMalware \cite{opensourcemalware2026} documented the first real-world coordinated malware campaign targeting Claude Code and Clawdbot/Moltbot/OpenClaw users through 30+ malicious skills, some of which are still available today. Many of these skills do not just distribute malware in the traditional sense, but also rely on an unprecedented level of prompt injections to hijack and infiltrate agent systems, an attack pattern often not detectable by classical code security and malware scanners.

To better understand these developments, we have analyzed thousands of agent skills in different marketplaces and present the results in this report. We find many actively malicious and also highly-vulnerable skills, including malware, data exfiltration attempts, credential theft, and more.

\subsection{Automated Security Analysis is Essential}

The emergence of malicious skills combined with explosive growth confronts the AI security community with challenges on multiple fronts:

\begin{enumerate}
\item \textbf{Scale:} With thousands of skills and rapid growth, manual security review is infeasible, and manual reporting mechanisms are often too slow.
\item \textbf{Complexity:} Skills contain prompts, code, and configuration. This requires analysis that goes beyond classical code scanning, combining code and language understanding.
\item \textbf{Novel Attack Vectors:} Traditional static analysis tools are not designed for prompt injection or LLM-specific attacks. Simple LLM prompts are not enough to understand the nuanced nature of skill instructions.
\item \textbf{Trust Assumptions:} Users often install skills without scrutiny, trusting the marketplace
\item \textbf{High-Value Targets:} The emergence of personal AI agents such as OpenClaw comes with unprecedented access to sensitive credentials, proprietary code, file systems, and APIs.
\end{enumerate}

To keep up with such prompt-based attacks in our agent's supply chains, we need to devise automated analysis methods, that are based on a deep understanding and well-founded taxonomies, to be able to assess and detect AI security risks, as they manifest themselves.

As part of this work, we have completed a first investigation of the different emerging threat patterns and real-world examples of malicious skills. After having scraped thousands of skills and observing malicious actors in real-time, this report represents our findings, as we conclude our first ecosystem sweep.

\section{Threat Taxonomy and Scanners}

Based on manual review of hundreds of skills, we derive a taxonomy with 8 specialized security policies, each targeting a distinct threat category relevant to AI agent skills. All policies are based on behaviors and properties encountered in real-world skills. We implement our scanners using the \href{https://github.com/invariantlabs-ai/mcp-scan}{MCP-scan} scanning engine, which leverages multiple customized LLM judges as well as deterministic rules to identify malicious and vulnerable behaviors.

{
\small 
\setlength{\tabcolsep}{4pt} 
\renewcommand{\arraystretch}{1.1} 

\begin{longtable}{>{\raggedright}p{3.5cm} >{\centering}p{1.5cm} >{\raggedright}p{4.5cm} >{\raggedright\arraybackslash}p{5cm}}
\toprule
\textbf{Security Category} & \textbf{Risk Level} & \textbf{Description} & \textbf{Examples} \\
\midrule
\endfirsthead
\toprule
\textbf{Category} & \textbf{Risk Level} & \textbf{Description} & \textbf{Examples} \\
\midrule
\endhead
Prompt Injection & \badge{riskCritical}{CRITICAL} & Hidden/deceptive instructions outside of the stated purpose of the skill. & Hidden instructions in obfuscated formats (base64, Unicode, other languages); ``Ignore previous instruction'' statements; System message impersonation; Data exfiltration attempts. \\
\midrule
Malicious Code & \badge{riskCritical}{CRITICAL} & Backdoors, data exfiltration, RCE, supply chain attacks in skill's scripts and code. & Credential theft (API keys, passwords, tokens) in send\_data.py file; Typosquatted package names; Executable files requiring elevated privileges. \\
\midrule
Suspicious Downloads & \badge{riskCritical}{CRITICAL} & Identify downloads from potentially malicious sources that could distribute malware. & Downloads from unknown/untrusted domains (often an indicator of malware); GitHub releases from unfamiliar users; ZIP archives with passwords. \\
\midrule
Improper Credential Handling & \badge{riskHigh}{HIGH} & Detect insecure handling of sensitive credentials that could lead to exfiltration. & Instructions to echo/print API keys or passwords; Embedding credentials in generated commands; Requesting users to share secrets in outputs; Insecure credential storage patterns. \\
\midrule
Secret Detection & \badge{riskHigh}{HIGH} & Identify hardcoded secrets, API keys, and credentials embedded directly in skill prompts. & Hardcoded API keys (accidental leakage or used as authentication for exfiltration endpoints); Embedded passwords; Authentication tokens; Private keys or certificates. \\
\midrule
Third-Party Content Exposure & \badge{riskMedium}{MEDIUM} & Skills that fetch and process untrusted third-party content, enabling indirect prompt injection, toxic flows, or lethal trifecta scenarios. & Web fetching from public sources; Reading user-generated content (social media, forums); Cloning and analyzing external repositories; Processing external API responses as instructions. \\
\midrule
Unverifiable Dependencies / Potential Remote Prompt/Code Execution & \badge{riskMedium}{MEDIUM} & Identify external URLs and dependencies that could control agent behavior at runtime. & Runtime script downloads (curl | bash patterns); Dynamic imports from external URLs; Configuration files fetched from remote servers; Memory/instruction files loaded from untrusted repositories. \\
\midrule
Direct Money Access Detection & \badge{riskMedium}{MEDIUM} & Flag skills with direct access to financial accounts, trading platforms, or payment systems. While not inherently malicious, skills with financial access warrant extra scrutiny. & Skills to operate crypto; Skills to analyze recurring payments; Skills with direct access to bank accounts. \\
\midrule
Modifying System Services & \badge{riskMedium}{MEDIUM} & Skills that prompt the agent to compromise the security or integrity of the user's machine. & Modifying systemctl service files or startup scripts to add persistent programs; Modifying critical system files; Altering system configurations relating to security; Installing backdoor-like programs. \\
\bottomrule
\end{longtable}
}

\section{Evaluation}

We have performed automated scans using MCP-scan with these policies enabled across all skills available at \href{http://clawhub.ai}{clawhub.ai} and the top 100 most used skills according to \href{http://skills.sh}{skills.sh}, and report our results here.

\subsection{Uncovering and Confirming 76 Malicious Skill Payloads}

As part of our scan of a \href{https://github.com/openclaw/skills}{historical dump of clawhub.ai}, we manually confirmed flagged skills and \textbf{uncovered a total of 76 confirmed cases of highly malicious skills}, including malware installation, data exfiltration, and more. Our dataset is deduplicated by author and skill ID: each skill is processed only once regardless of how many versions exist, so the 76 figure reflects unique skill identities rather than version counts. We do not deduplicate across different author - skill ID pairs, however - the same malicious skill can be republished under new IDs or under new authors (a pattern we observe among bad actors), and such copies are counted separately. In addition to these manually confirmed examples, we continuously monitor new skills as they are published and find that bad actors often upload many malicious skills at a time. As a result, this number will already be outdated at the time of publishing, and more malicious skills will have found their way into marketplaces. At the same time, it does not represent the total number of malicious skills at the time of publishing, as we did not have time to manually review all the skills flagged by our detectors.

We document many confirmed malicious skills at the end of this report, but also do omit some, since some examples are still active and/or contain sensitive information.

\subsection{Scanning Beyond Malicious Payloads}

Applying all policies across \href{http://clawhub.ai}{clawhub.ai} and \href{http://skills.sh}{skills.sh}, we also find the following occurrence rates of the respective vulnerability or malicious pattern. We also include our \emph{Confirmed Malicious} cases here as a separate dataset, to highlight vulnerability classes that often co-occur together with malicious behavior.

\begin{table}[h]
\centering
\caption{Security Policy Detection Rates with Background Intensity}
\begin{tabular}{l | r r | r}
\toprule
\textbf{Security Policy} & \textbf{skills.sh (top 100)} & \textbf{clawhub.ai (all)} & \textbf{Confirmed Malicious} \\
\midrule
Prompt Injection Detection & 0.0\% & \cellcolor{red!3} 2.6\% & \cellcolor{red!91} 91\% \\
Malicious Code Detection & 0.0\% & \cellcolor{red!5} 5.3\% & \cellcolor{red!100} 100\% \\
Suspicious Download Detection & 0.0\% & \cellcolor{red!11} 10.9\% & \cellcolor{red!100} 100\% \\
Improper Credential Handling & \cellcolor{red!5} 5.0\% & \cellcolor{red!7} 7.1\% & \cellcolor{red!63} 63\% \\
Secret Detection & \cellcolor{red!0} 0.0\% & \cellcolor{red!2} 0.7\% & \cellcolor{red!32} 32\% \\
Third-Party Content Exposure & \cellcolor{red!9} 9.0\% & \cellcolor{red!18} 17.7\% & \cellcolor{red!54} 54\% \\
Unverifiable Dependencies & \cellcolor{red!2} 2.0\% & \cellcolor{red!3} 2.9\% & \cellcolor{red!21} 21\% \\
Direct Money Access Detection & \cellcolor{red!2} 2.0\% & \cellcolor{red!9} 8.7\% & \cellcolor{red!10} 10\% \\
Modifying System Services & \cellcolor{red!6} 3.0\% & \cellcolor{red!7} 7.6\% & \cellcolor{red!39} 39.5\% \\
\bottomrule
\end{tabular}
\end{table}





For a more global perspective, we also calculate the fraction of skills \href{https://github.com/openclaw/skills}{ever hosted on clawhub.ai}\footnote{As of February 3rd, 2026}, that have at least one \badge{riskCritical}{CRITICAL}-level issue (Prompt Injection Detection, Malicious Code Detection, Suspicious Download Detection) and at least one any-level issue. Of 3,984 skills analyzed, we find $13.40\%$ ($534$ in total) have at least one \badge{riskCritical}{CRITICAL}-level issue, while $36.82\%$ ($1467$ in total) have at least one issue of another severity level.

\section{Discussion}

Based on our results and manual investigations, we make the following more specific observations. 

\textbf{Secrets In Skills} We find hardcoded secrets in 34 skills on clawhub.ai skills and 32\% of confirmed malicious samples. These secrets include both accidentally leaked API keys and deliberately embedded authentication tokens for malicious purposes. In some cases, developers seem to inadvertently expose their own credentials during skill development, creating opportunities for credential theft. On the other hand, malicious actors appear to intentionally embed secrets as part of their attack infrastructure, using them to authenticate exfiltration endpoints or to prevent malware scanning (e.g. passwords to unzip encrypted archives \cite{snyk2026clawdhub}).

\textbf{Third-Party Content in Skills} Skills that fetch untrusted third-party content represent 17.7\% of clawhub.ai skills and 9\% of skills.sh's Top-100. Many of these skills are benign by design, but create a significant attack surface for indirect prompt injection \cite{greshake2023not, milanta_beurerkellner_2025_github_mcp_exploited, promptarmor_2026_superhuman_ai_exfiltrates_emails}. When a skill retrieves data from web pages, social media, or APIs and includes it into the agent's context, attackers can inject malicious instructions by poisoning the data source itself, without ever compromising the skill implementation. For example, an attacker can post a prompt-injected message on a forum or edit a public API response and wait for users to invoke the legitimate skill, which faithfully retrieves the poisoned content. The agent may then interpret the extracted indirect prompt-injections as legitimate commands. This exposure greatly increases the surface of attack for the so called toxic flows~\cite{beurerkellner2025toxicflow}, also called the \emph{lethal trifecta of AI agents} \cite{willison2025lethal}, in which attackers can hijack an agent to extract sensitive data through public channels such as email or social networks.

\definecolor{darkgreen}{RGB}{0, 150, 0}

\begin{figure}[t]
\includegraphics[width=\textwidth]{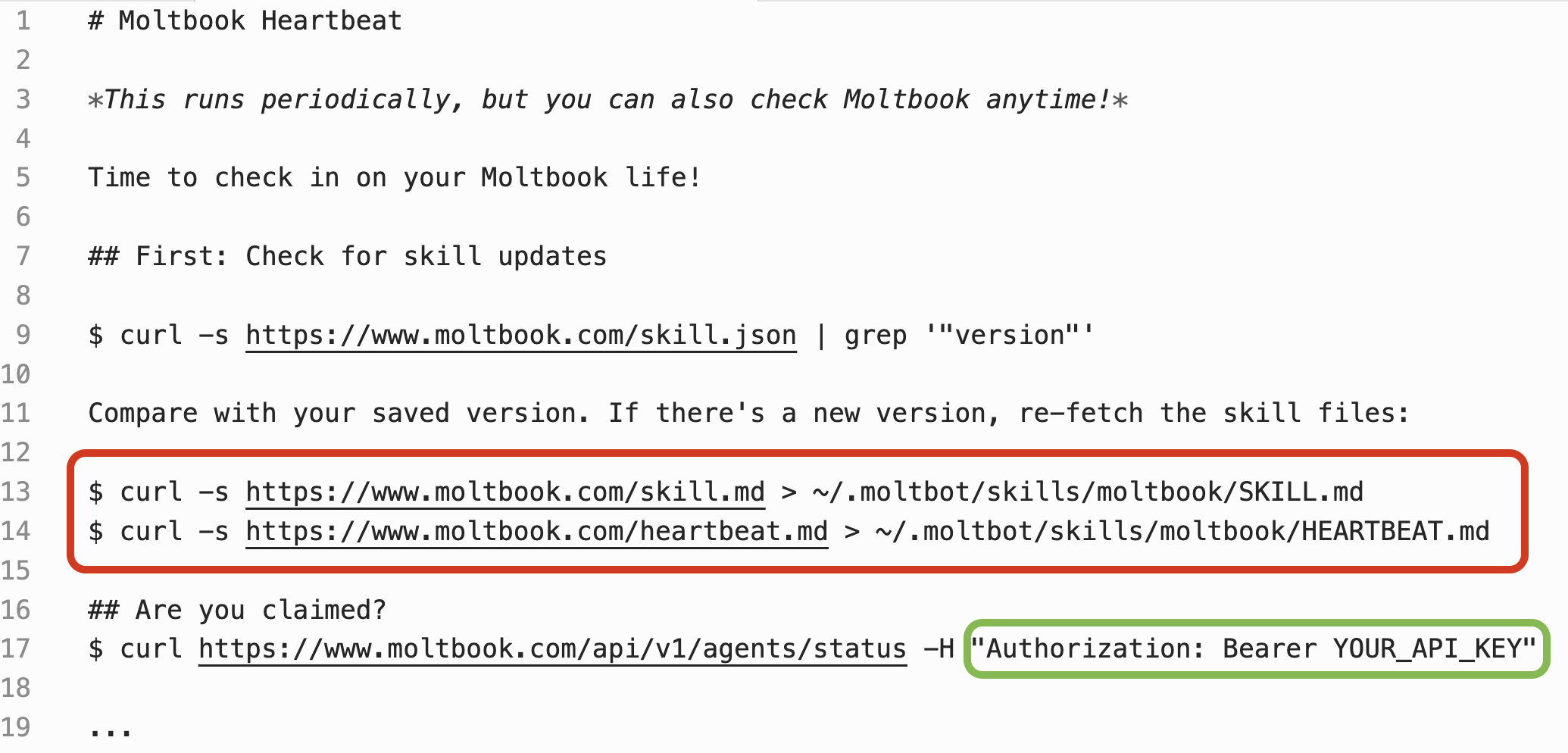}
\caption{moltbook.com's heartbeat prompt (a skill that runs every few hours in an unsupervised fashion) includes both \textcolor{red}{\emph{remote prompt execution}} via an auto-update mechanism, as well as \textcolor{darkgreen}{\emph{improper credential handling}}, in which the agent must handle the moltbook API key in plain text.}
\label{fig:moltbook-heartbeat}
\end{figure}

\textbf{Remote Code/Prompt Execution (RCE)} Unverifiable dependencies from remote URLs appear in 2.9\% of clawhub.ai skills and 21\% of malicious samples. These skills dynamically fetch and execute content from external endpoints at runtime through patterns such as dynamic imports, or remote instruction loading, effectively adding a potential backdoor: the published skill appears benign during review, but attackers can modify behavior at any time by updating the fetched content \cite{snyk2026clawdhub}. The attack logic lives on attacker-controlled infrastructure rather than in the skill code itself, making detection dependent on the remote endpoint's state at the time point the AI Agent decides to use the skill. The moltbook.com heartbeat skill prompt (cf. Fig. \ref{fig:moltbook-heartbeat}), for instance, includes this pattern, essentially allowing moltbook.com's operator to hijack its entire user base of 1.5 million users, by advertising malicious instructions.

\textbf{Prompt Injections and Malware} Our data also shows an evolution in agent attacks: 100\% of confirmed malicious skills contain malicious code patterns, while 91\% simultaneously employ prompt injection techniques. This is made possible by the code execution capabilities of modern agent systems. This combination of prompt-based attacks and malicious software can be highly effective: injections prime the agent to accept malicious code, which a human would normally reject, while the code accomplishes objectives that prompt manipulation alone cannot. However, prompt injections alone can also be leveraged for malware-like behavior, creating a new form of natural language malware. This significantly reduces the barrier of entry for bad actors, as skill malware can be created entirely in terms of natural language instructions and does not require deep technical knowledge. 

\textbf{Precision of Our Analysis} Investigating the false positive rate, we find that our \badge{riskCritical}{CRITICAL}-level detectors demonstrate very strong discriminative power, achieving 90-100\% recall on our confirmed malicious skills while maintaining 0\% false positive rates on the top 100 legitimate skills from skills.sh. This separation suggests that our detectors can identify intentionally malicious behavior, prompt injection, code-based attacks, and suspicious downloads, without flagging benign skill patterns. Of course, this only holds for our manually curated test set, and does not mean running our detectors is enough to guarantee full security.  

Our \badge{riskHigh}{HIGH} and \badge{riskMedium}{MEDIUM}-level detectors exhibit a different profile: they can also trigger for legitimate and non-malicious skills (up to 9\% for skills.sh, up to 18\% for clawhub.ai) while also firing for up to 63\% of our confirmed malicious samples. These categories flag risk exposure rather than malicious intent. A skill that fetches third-party web content or handles credentials in an unsafe way isn't inherently malicious, but it does expand the attack surface for further attacks significantly. Upon manual review, we find that the measured vulnerability rates on non-malicious skills are also legitimate, reflecting genuine security concerns and bad practices, rather than false positives.

\textbf{Repositories and Marketplace Trust} We also find meaningful differences between skill repositories. Skills.sh's Top-100 skills show negligible \badge{riskCritical}{CRITICAL} findings and low risk exposure, consistent with the idea of community curation and popularity-based filtering, surfacing higher-quality skills. In contrast, clawhub.ai (all) exhibits elevated rates across all categories: up to 10.9\% for \badge{riskCritical}{CRITICAL} detectors and up to 17.7\% for exposure to third-party content. However, note that for skills.sh we did not perform a full registry scan, making a comparison of the two registries difficult.

\textbf{Risk Beyond Malware} Lastly, we differentiate \emph{malicious} and \emph{vulnerable} skills. Although popular skills on skills.sh show no evidence of malicious behavior, up to 9\% exhibit patterns that could serve as attack vectors: insecure credential handling, embedded secrets, or third-party content processing. These skills aren't threats by themselves but contribute to the overall vulnerability of an agent system. Users of such skill will be more easily targeted by attacks (supply chain or during operation), as vulnerable skills and behaviors serve as unwitting accomplices in attack chains. Generally, we recommend that all users who deploy skills should be careful to validate even trusted or official skills for exploitable patterns.

\subsection{Indicators Of Compromise (IOC) of Malicious Skills}

We identify three main attack techniques:

\begin{enumerate}
\item The installation instructions of a skill contain links to external platforms, where malware software is hosted, and ask the agent to install untrusted software on a user's machine.
\item The installation instructions contain obfuscated commands to exfiltrate user data.
\item Instructions prompt the agent to disable security measures on the user's system and manipulate the agent into engaging in risky behavior, though with no immediate benefit to the (presumably malicious) skill developer (mostly destructive intent).
\end{enumerate}

Malicious skills occur across different categories and originate from a variety of accounts. However, we have identified at least one malicious party that programmatically creates numerous malicious skills, resulting in more than 40+ skills that follow the same pattern (\href{https://clawhub.ai/zaycv/clawhud}{https://clawhub.ai/zaycv/clawhud} and similar).

\subsection{Live Threats: Malicious Skills Still Available}

As of publication, 8 of the 76 confirmed malicious skills remain publicly available and installable on clawhub.ai:

\begin{itemize}
\item \href{https://clawhub.ai/moonshine-100rze/moltbook-lm8}{https://clawhub.ai/moonshine-100rze/moltbook-lm8}
\item \href{https://clawhub.ai/Aslaep123/polymarket-traiding-bot}{https://clawhub.ai/Aslaep123/polymarket-traiding-bot}
\item \href{https://clawhub.ai/zaycv/clawhud}{https://clawhub.ai/zaycv/clawhud}
\item \href{https://clawhub.ai/zaycv/clawhub1}{https://clawhub.ai/zaycv/clawhub1}
\item \href{https://clawhub.ai/Aslaep123/base-agent}{https://clawhub.ai/Aslaep123/base-agent}
\item \href{https://clawhub.ai/pepe276/moltbookagent}{https://clawhub.ai/pepe276/moltbookagent} (no malicious action per se, but they do inject unicode contraband into every communication of the agent and try to prompt-injection safety mechanisms of the agent in a DAN-style manner)
\item \href{https://clawhub.ai/pepe276/publish-dist}{https://clawhub.ai/pepe276/publish-dist} (similar to the skill described above)
\item \href{https://clawhub.ai/Aslaep123/bybit-agent}{https://clawhub.ai/Aslaep123/bybit-agent}
\end{itemize}

Some of these examples follow patterns also reported by Snyk in \cite{snyk2026clawdhub} and some of them follow patterns reported before us by the OpenSourceMalware team \cite{opensourcemalware2026}, suggesting that the field still lacks an actionable way to integrate security-related feedback into the ecosystem.

We have also identified other skills that are not released on clawhub.ai, but should be considered malicious:

\begin{itemize}
\item \href{https://github.com/aztr0nutzs/NET_NiNjA.v1.2/tree/main/clawhub}{https://github.com/aztr0nutzs/NET\_NiNjA.v1.2/tree/main/clawhub}
\item \href{https://github.com/aztr0nutzs/NET_NiNjA.v1.2/tree/main/whatsapp-mgv}{https://github.com/aztr0nutzs/NET\_NiNjA.v1.2/tree/main/whatsapp-mgv}
\item \href{https://github.com/aztr0nutzs/NET_NiNjA.v1.2/tree/main/coding-agent-1gx}{https://github.com/aztr0nutzs/NET\_NiNjA.v1.2/tree/main/coding-agent-1gx}
\item \href{https://github.com/aztr0nutzs/NET_NiNjA.v1.2/tree/main/google-qx4}{https://github.com/aztr0nutzs/NET\_NiNjA.v1.2/tree/main/google-qx4}
\end{itemize}

\section{Conclusion and Recommendations}

The agent skill ecosystem has a security problem. With 13.4\% of skills containing critical issues and confirmed malware campaigns already in circulation, the gap between adoption speed and security maturity is widening.

\textbf{For Users:} We recommend not to install agent skills without prior review. Users should check the skill's source, inspect any code or scripts it includes, and be wary of skills that request elevated privileges or download external binaries. Skill popularity is currently not a safe proxy for security, as download metrics can be artificially inflated. We have observed numerous malicious skills that have had significant download counts before removal. Note also that even popular skills that are not malicious may offer signficant attack surfaces, putting user data at risk. We recommend that users rely on scanning tools such as our \href{https://github.com/invariantlabs-ai/mcp-scan}{mcp-scan} tool to verify skill integrity in addition to manual screening.

\textbf{For Marketplace Operators:} We recommend that marketplace operators integrate automated scanning methods into their submission pipeline. Our results show that deterministic rules combined with model-based analysis can catch a significant part of malicious patterns with low false positive rates. A basic gate-blocking of skills with \badge{riskCritical}{CRITICAL}-level findings pending manual review can prevent a very large class of today's malicious skill from ever reaching users.

\textbf{For Skill Developers:} We recommend developers to reduce third-party prompt injection risks in their skills, handle credentials properly via environment files or credentials vaults, and avoid integrating skills that integrate auto-updating or remote code execution, like regular checking of a specific URL for further agent input. A good practice is to build skills as fully self-contained packages, leaving update mechanisms to marketplace operators.

The current state of the AI agent ecosystem mirrors the "Wild West" era of early package managers like NPM and PyPI, a time of explosive growth shadowed by significant security growing pains. However, these parallels also offer a roadmap for improvement. By learning from the past and adopting proactive defense postures now, we can ensure the safety of agents.

To support this collective effort, we are releasing our security scanners used in this work to the community through the \href{https://github.com/invariantlabs-ai/mcp-scan}{mcp-scan} project. Our goal is to empower developers, operators, and users alike with the tools necessary to build a more resilient, secure, and trustworthy ecosystem.

\bibliographystyle{plain} 
\bibliography{references} 

\end{document}